\documentclass[aps,onecolumn,superscriptaddress,preprint]{revtex4}
\usepackage{graphicx} 
\usepackage{amssymb}
\usepackage{dcolumn}
\usepackage{bm} 
\newcommand{\dg}{d_{\mathrm{LAO}}}
\newcommand{\Vg}{V_{\mathrm{GS}}}
\newcommand{\Igs}{I_{\mathrm{G}}}
\newcommand{\Ids}{I_{\mathrm{D}}}
\newcommand{\Vth}{V_{\mathrm{Th}}}
\newcommand{\Vds}{V_{\mathrm{DS}}}
\newcommand{\Vp}{V_{\mathrm{P}}}
\newcommand{\gd}{g_{\mathrm{D}}}

\begin{document}
\title{Transistor operation and mobility enhancement in top-gated LaAlO$_3$ / SrTiO$_3$ heterostructures}
\author{Masayuki Hosoda}
\affiliation{Stanford Institute for Materials and Energy Sciences, SLAC National Accelerator Laboratory, Menlo Park, CA 94025, USA}
\affiliation{Department of Advanced Materials Science, The University of Tokyo, Kashiwa, Chiba 277-8561, Japan}
\author{Yasuyuki Hikita}
\affiliation{Stanford Institute for Materials and Energy Sciences, SLAC National Accelerator Laboratory, Menlo Park, CA 94025, USA}
\author{Harold Y. Hwang}
\affiliation{Stanford Institute for Materials and Energy Sciences, SLAC National Accelerator Laboratory, Menlo Park, CA 94025, USA}
\affiliation{Geballe Laboratory for Advanced Materials, Department of Applied Physics, Stanford University, Stanford, CA 94305, USA}
\author{Christopher Bell}
\email{cbell@slac.stanford.edu}
\affiliation{Stanford Institute for Materials and Energy Sciences, SLAC National Accelerator Laboratory, Menlo Park, CA 94025, USA}
\date{\today}
\begin{abstract}
We report the operation of LaAlO$_3$ / SrTiO$_3$ depletion mode top-gated junction field-effect transistors using a range of LaAlO$_3$ thicknesses as the top gate insulator. Gated Hall bars show near ideal transistor characteristics at room temperature with on-off ratios greater than 1000. Lower temperature measurements demonstrate a systematic increase in the Hall mobility as the sheet carrier density in the channel is depleted via the top gate, providing a route to higher mobility, lower density electron gases in this system.
\end{abstract}
\maketitle

\newpage

The electric field-effect has been successfully applied in a top-gate geometry to create normally-off transistors in undoped SrTiO$_3$ (STO).\cite{shibuya_apl2004,nakamura_apl2006} Diverse properties have been studied including superconductivity,\cite{ueno_natmat2008,leey_prl2011} thermo-electricity,\cite{ohta_natcomm2010} Kondo physics,\cite{leem_prl2011} and percolation effects.\cite{li_prl2012} The normally-on interface between LaAlO$_3$ (LAO) and TiO$_2$ terminated (100) STO \cite{ohtomo_nature2004} has also been electrostatically modulated from the back-side using the STO substrate as the gate dielectric.\cite{thiel_science2006,caviglia_nature2008,BellPRL2009,BenShalomPRL2010SOC,CavigliaPRL2010Rashba,joshua_natcomm2012,BertPRB2012,FleskerPRB2012,KaliskyNcomm2012} Although back-gating is rather robust and reliable, it simultaneously changes many parameters, most notably the confining electric field strength at the interface, Hall mobility and sheet carrier density.\cite{BellPRL2009} Hence back-gating by itself does not allow low sheet carrier densities to be reached while maintaining relatively high Hall mobilities. Nor can the Rashba spin-orbit coupling be controlled independently of the sheet carrier density. Top-gating is a natural complimentary technique to tune the electronic properties: as is well-known in conventional semiconductor systems both the top- and back-gate geometries change the confining potential of the electron gas, but the spatial extent of the envelope electron wavefunction is altered in quite distinct ways for the two cases.\cite{hirakawa_prl1985} 

The challenges of switching on devices at low temperatures with a top gate in STO,\cite{nakamura_apl2006} can be mitigated by using the depletion mode of junction gate field-effect transistors. Somewhat surprisingly however, given the attractively large band-gap of the over-layer in LAO/STO heterostructures, top-gate depletion devices have only recently been demonstrated by F\"org {\it et al.}\cite{ForgAPL2012} over a temperature range 173 K $\le T\le$ 373 K. These devices had a fixed LAO thickness, $\dg = 9 $ unit cells (uc), and short (20 $\mu$m - 200 $\mu$m) and wide (1600 $\mu$m) channels, to enable device operation with a reduced channel resistance and gate leakage. However the two-probe device geometry prevented the determination of the sheet carrier density, $n$, and hence the Hall mobility. Here we employed micrometer-scale Hall bars to enable precise Hall measurements, and study the field effect on the carrier density, for various $\dg$. Ideal transistor operation with top-gating was demonstrated, as evidenced by the direct scaling of the mobile sheet carrier density obtained from the Hall effect with the expected modulation by the top-gate voltage, $\Vg$. A sheet carrier density modulation of up to $1.7\times10^{13}$ cm$^{-2}$ was achieved with $\Vg$ in the range -1 V $\le \Vg \le$ +1 V and 2 K $\le T \le$ 300 K. Notably, at lower temperatures the Hall mobility was significantly enhanced with field effect depletion of the electron gas, demonstrating the potential of top-gating to reach low sheet carrier density and high mobilities. 

\begin{figure}[h]
\includegraphics[width=8cm,clip]{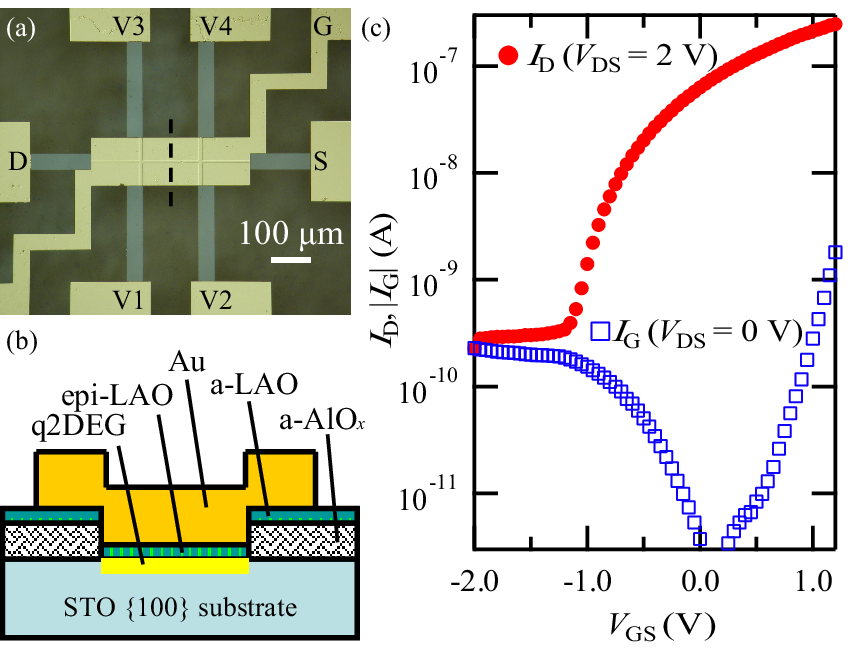}
\caption{(color online) (a) Optical microscope image of a device with central channel dimensions 5 $\mu$m $\times$ 150 $\mu$m between the centers of the voltage contacts (V1, V2) and (V3, V4). Source, gate and drain are labeled S, G and D respectively. (b) Schematic cross-section of the device (not to scale) taken along the dashed line in (a). ``a'' and ``epi'' refer to amorphous and epitaxial layers in the cross-section, and q2DEG is the electron gas. The amorphous AlO$_x$ thickness is $\sim 50$ nm. (c) $\Ids-\Vg$ and $\Igs-\Vg$ characteristics for a typical device. Here $\dg = 16$ uc, and $T = 300$ K.}\label{fig1}
\end{figure}

The LAO/STO heterostructures were fabricated by pulsed laser deposition on TiO$_2$ terminated STO (100) substrates using a pre-deposited amorphous AlO$_x$ hard mask to define the Hall bar as described elsewhere.\cite{BellPRL2009} The LAO thicknesses were in the range 4 uc $\le \dg \le$ 22 uc. The channel length covered by the gate was 400 $\mu$m, and the channel width in the range $W = 5-200$ $\mu$m. The top electrode was formed using {\it ex-situ} sputtered Au with thickness $\sim 100$ nm, and the source and drain were contacted with Al ultrasonic wirebonding. Figure \ref{fig1}(a) and (b) shows an optical image of a typical device, showing the source, drain and gate contacts, together with a schematic device cross-section. All transport measurements were carried out with a semiconductor parameter analyzer in DC mode, and capacitance measurements were made using a LCR meter. When the devices were on, the gate leakage current $\Igs$ was always significantly smaller than the drain current $\Ids$, an example of which is shown by the $I-V$ characteristics in Fig$.$ \ref{fig1}(c). The gate leakage shows a similar form to previous studies.\cite{singh_natphys2011} 

\begin{figure}[h]
\includegraphics[width=8cm,clip]{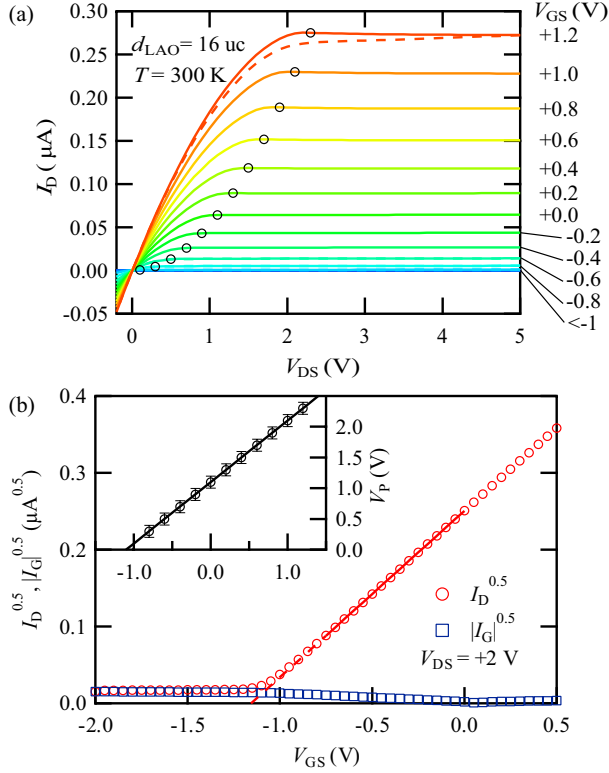}
\caption{(color online) (a) Transistor operation of a $\dg = 16$ uc device at room temperature, for various $\Vg$ taken in 0.2 V steps. All data are taken with increasing $\Vds$ except for the dashed line at $\Vg=1.2$ V, which is for decreasing $\Vds$. Circles mark the co-ordinates $(\Vp,I_\mathrm{D,sat})$ on the respective traces. $W = 100$ $\mu$m. (b) $\sqrt{\Ids}$-$\Vg$ plot, showing linear behavior in the saturation region. The abscissa intercept of the linear fit gives $\Vth = -1.15 \pm 0.05$ V for this device, where the error bar is associated with the uncertainty given by the finite gate leakage. Inset: $\Vp-\Vg$ (open circles) obtained from the $\gd-\Vg$ intercepts. Line is a best fit to the form $\Vp = \Vg + \Vth$, where $\Vth$ is the fit parameter.}\label{fig2}
\end{figure}

Clear normally-on transistor characteristics at $T = 300$ K are shown in Fig$.$ \ref{fig2} for the case of $\dg = 16$ uc. Pinch-off is observed for -0.8 V $\le \Vg \le$ +1.2 V, with a clear saturation at higher $\Vds$. Differentiating these data to obtain the channel conductance $\gd = \frac{\mathrm{d}\Ids}{\mathrm{d}\Vds}$, and extrapolating to zero $\gd$, we find the pinch-off voltage $\Vp$ as a function of $\Vg$, as shown in the inset of Fig$.$ \ref{fig2}(b). The corresponding current at $\Vds = \Vp$, $I_\mathrm{D,sat}$, defines the co-ordinates $(\Vp,I_\mathrm{D,sat})$ for a fixed $\Vg$, as shown by the circles in Fig$.$ \ref{fig2}(a). The threshold voltage could be obtained as $\Vth = -1.15 \pm 0.05$ V by fitting the $\Vp-\Vg$ data to the relationship $\Vp = \Vg + \Vth$.\cite{sze_book} Taking a transfer curve at a source-drain voltage of $\Vds = 2$ V, as shown in Fig$.$ \ref{fig2}(b), essentially ideal quadratic form of $\Ids-\Vg$ is found, as expected from the gradual-channel model in the saturation regime.\cite{sze_book} These data allow a clear extrapolation of the threshold voltage as shown, giving $\Vth = -1.1 \pm 0.1$ V. The good agreement between these two $\Vth$ values, obtained by two different analysis methods, suggests that the mobile carriers in the channel are ideally tuned by $\Vg$, with on-off current ratio's of greater than 1000 being achievable. Small hysteresis could be observed between increasing and decreasing $\Vds$ in the transfer characteristics, as shown for the case of $\Vg = +1.2$ V in Fig$.$ \ref{fig2}(a), where the difference in $\Ids$ between the increasing and decreasing sweep of $\Vds$ is of the scale of 10 nA.

\begin{figure}[h]
\includegraphics[width=8cm,clip]{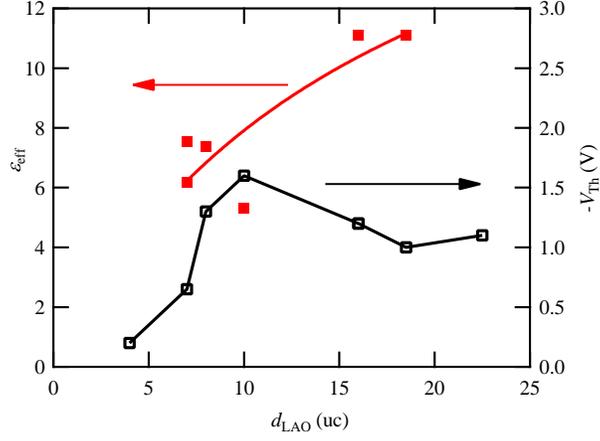}
\caption{(color online) Scaling of $\Vth$ (right axis) at $T = 300$ K and $\varepsilon_{\mathrm{eff}}$ at $T \sim 100$ K (left axis) with $\dg$. Lines between $\Vth$ data are guides to the eye. For the $\varepsilon_{\mathrm{eff}}$ data the line is a best fit to a series dead layer model giving $\varepsilon_{\mathrm{bulk}}= 21.6$, $\varepsilon_{\mathrm{dead}}= 4.3$ and $d_{\mathrm{dead}} = 4.3$ uc. $W = 5$ $\mu$m for all devices.}\label{fig3}
\end{figure}

We systematically varied $\dg$, and were able to achieve transistor operation over the thickness range 4 uc $\le \dg \le$ 22 uc. The scaling of $\Vth(\dg)$ is shown in Fig$.$ \ref{fig3}. $\Vth$ does not scale monotonically with $\dg$, as would be the case for constant sheet carrier density, via the relation $\Vth = - e\dg n(\Vg = 0) / \varepsilon_0 \varepsilon_r$, here $e$ is the electronic charge, $\varepsilon_0$ and $\varepsilon_r$ are the vacuum and LAO relative permittivities respectively, and the sign convention is taken such that $n>0$ and $\Vth < 0$. It is notable that a systematic variation of the transport properties has been seen in this thickness regime elsewhere.\cite{BellAPL2009,CancellieriEPL2010,singh_natphys2011} In order to more carefully examine this issue, we characterized several samples at lower temperatures where the Hall effect could be more reliably measured to directly measure the sheet carrier density change, and relate it to the capacitive character of the gate dielectric, since $\Delta n / \Delta \Vg = C_{\mathrm{eff}}$. Here the effective capacitance $C_{\mathrm{eff}}$ is modeled as a `dead' layer with thickness $d_{\mathrm{dead}}$ and dielectric constant $\varepsilon_{\mathrm{dead}}$ in series with a bulk-like LAO layer with dielectric constant $\varepsilon_{\mathrm{bulk}}$. From a simple series capacitor model $\dg/\varepsilon_{\mathrm{eff}} = (\dg - d_{\mathrm{dead}})/\varepsilon_{\mathrm{bulk}} + d_{\mathrm{dead}}/\varepsilon_{\mathrm{dead}}$. We note that the dead layer thickness also includes possible suppression of the dielectric properties at the Au/LAO interface.\cite{spaldin_nature} Assuming otherwise ideal capacitor behavior, $\Delta n / \Delta \Vg$ for various $\dg$ can be used to extract $\varepsilon_{\mathrm{eff}}(\dg)$, and estimate $\varepsilon_{\mathrm{dead}}$, $\varepsilon_{\mathrm{bulk}}$ and $d_{\mathrm{dead}}$. The measured $\varepsilon_{\mathrm{eff}}(\dg)$ is shown in Fig$.$ \ref{fig3}, together with a theoretical fit which gives $\varepsilon_{\mathrm{dead}} \sim 4.3$, $\varepsilon_{\mathrm{bulk}} \sim 21.6$ and $d_{\mathrm{dead}} \sim 4.3$ uc. These estimates, in particular the value of $\varepsilon_{\mathrm{bulk}}$ emphasize that the atomic scale nature of interface between the gate and the electron gas.         
      
\begin{figure}[h]
\includegraphics[width=8cm,clip]{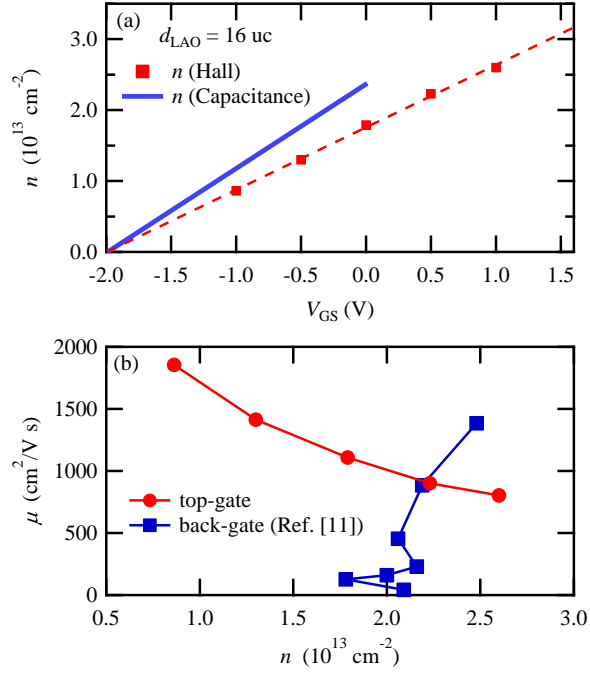}
\caption{(color online) (a) Control of the sheet carrier density $n$ with $\Vg$ at $T = 2$ K. Solid line shows the ideal predicted rate of carrier density tuning based on the capacitance measurements at $T \sim 150$ K, taking $n = 0$ cm$^{-2}$ at $\Vg = \Vth = -2$ V. (b) Hall mobility enhancement with top-gate depletion of $n$ at $T = 2$ K. Back-gate mobility suppression data from Ref$.$ \cite{BellPRL2009}. LAO thickness, $\dg= 16$ uc, $W = 5$ $\mu$m.}\label{fig4}
\end{figure}

With the establishment of robust transistor operation at room temperature, it is important to consider if the top-gating geometry can be successfully applied at lower temperatures. Figure \ref{fig4}(a) shows clearly that $n$ can be ideally tuned with the top gate in a $\dg = 16$ uc, $W = 5$ $\mu$m sample at $T = 2$ K, where we compare the $n$ variation from the Hall effect to that predicted from the measured device capacitance (940 pF with top-gated channel area 0.0504 mm$^2$) at $\Vg = 0$ V (solid line), assuming that $n = 0$ cm$^{-2}$ at $\Vg = \Vth = -2$ V. Here the capacitance was measured in a frequency range of 20-100 Hz with an AC modulation voltage of 50 mV. In this range the loss tangent was $< 0.04$ and the gate leak conduction is negligibly small. In addition to the good agreement between the capacitance data and the Hall effect, as $n$ was depleted the Hall mobility $\mu$ showed a clear increase: the opposite tendency compared to back-gate field effect modulation,\cite{BellPRL2009} as shown in Fig$.$ \ref{fig4}(b).

The dramatic asymmetry in the mobility modulation between back- and top-gating can be understood by considering the action of the gate voltages on the confined electron gas. For negative back-gating the applied electric field tends to confine the electrons closer to the interface where they are scattered more strongly.\cite{BellPRL2009} In the top gate geometry by contrast, the center of weight of the envelope wavefunction of the electron gas is pushed away from the interface as the electron gas is depleted and the self-consistent confining electric field is reduced.\cite{hirakawa_prl1985} This results in a local three-dimensional electron density that becomes smaller, and located in the less disordered bulk STO as $\Vg \rightarrow \Vth$, producing the observed enhancement of the Hall mobility as $n$ is decreased. The significance here is that the systematic $n$ suppression achieved by the top gate voltage can provide access to a regime with both smaller carrier density and less disorder (higher mobility), where other methods have experienced difficulties in approaching this regime. By combining the top-gate with an additional back-gate at low temperatures, we can also envisage independently tuning the sheet carrier density and Hall mobility to control the electronic character of this fascinating system.

The authors acknowledge support by the Department of Energy, Office of Basic Energy Sciences, Materials Sciences and Engineering Division, under contract DE-AC02-76SF00515. M. H. acknowledges partial support from the ONR-MURI number N00014-12-1-0976.

\newpage

\newpage

\newpage

\newpage

\end{document}